\documentclass[3p,times,procedia]{elsarticle}
\usepackage{nupha_ecrc}

\volume{00}

\firstpage{1}

\journalname{Nuclear Physics A}

\runauth{}

\jid{nupha}

\jnltitlelogo{Nuclear Physics A}

\usepackage{amssymb}

\usepackage{lineno}

\usepackage[figuresright]{rotating}
\usepackage{xspace}

\RequirePackage{mhchem}

\newcommand*{\pp}{\ensuremath{pp}\xspace}

\newcommand*{\PbPb}{\ce{Pb}+\ce{Pb}\xspace}

\newcommand*{\sqn}{\ensuremath{\sqrt{s_{_\text{NN}}}}\xspace}
\newcommand*{\sqs}{\ensuremath{\sqrt{s}}\xspace}
\newcommand{\lns}{\ensuremath{\ln(\kern -0.2em\sqrt{s})}\xspace}

\newcommand*{\RAA}{\ensuremath{R_{\ce{AA}}}\xspace}

\newcommand*{\vt}{\ensuremath{v_{2}}\xspace}

\newcommand*{\pT}{\ensuremath{p_{\ce{T}}}\xspace}

\newcommand*{\pID}{\ensuremath{p_{\ce{ID}}}\xspace}
\newcommand*{\Deltaploss}{\ensuremath{\Delta p}\xspace}

\let\psii=\psi  
\renewcommand*{\psi}{\ensuremath{\psii}\xspace}

\newcommand*{\Jpsi}{\ensuremath{J/\psi}\xspace}

\newcommand{\papertype}{proceeding}

\begin{document}

\begin{frontmatter}


\title{Title\tnoteref{label1}}
\tnotetext[label1]{}
\ead{alexander.milov@weizmann.ac.il}

\dochead{}

\title{The \RAA and \vt of muons from heavy-quark decays in \PbPb collisions at \sqn=2.76~TeV with the ATLAS detector}


\author{Alexander Milov for the ATLAS Collaboration}
\address{Weizmann Institute of Science, 234 Herzl Street, Rehovot 7610001 Israel}

\begin{abstract}

The ATLAS experiment measures the production of muons coming from the decays of heavy flavour particles in the kinematic interval $4 {<} \pT {<} 14$~GeV and $|\eta|{<}1$. The measurement is performed in $\sqs = 2.76$~TeV \pp collisions and over the centrality range of (0--60)\% in $\sqn = 2.76$~TeV \PbPb collisions. The heavy flavour muon differential cross-sections and per-event yields are measured in \pp and \PbPb collisions, respectively. The nuclear modification factor measured in 0--10\% most central collisions is observed to be approximately equal to 0.4 and independent of \pT within uncertainties, which indicates suppressed production of heavy flavour muons in \PbPb collisions. The muon yields are also measured as a function of the azimuthal angle with respect to the event plane. Fourier coefficients associated with the second harmonic modulation vary slowly with \pT and show a systematic variation with centrality that is characteristic of other elliptic anisotropy measurements.
\end{abstract}


\end{frontmatter}


\section{Introduction}
\label{sec:intro}

Heavy quarks provide an important probe of the properties of the quark-gluon plasma created in high-energy nuclear-nuclear collisions \cite{vanHees:2004gq,
Herzog:2006gh,vanHees:2007me,Horowitz:2007su,Uphoff:2011ad,He:2011qa,Cao:2013ita}. 
Heavy quarks , with the mass exceeding the temperature of the surrounding plasma $T{\sim} 200-500$~MeV~\cite{Abreu:2007kv} are mostly produced early in the collision. Their production rates that can be calculated using pQCD, and their subsequent interactions reflect themselves in  experimentally observable signatures. At high transverse momenta, greater than the quark mass, heavy quarks are thought to lose energy similar to light quarks but with mass-dependent modifications to the pattern of collisional and radiative energy loss \cite{Djordjevic:2003zk,Gossiaux:2010yx,Gossiaux:2009hr}. At lower transverse momenta the quarks are thought to diffuse in the plasma losing energy and partially thermalizing \cite{Moore:2004tg}. Interacting with the medium heavy quarks may acquire an azimuthal anisotropy due to the collective expansion of the medium. Past measurements of heavy flavour quarks at RHIC and the LHC using semi-leptonic decays \cite{Adare:2006nq, Adare:2010de, Abelev:2012qh, Adam:2015pga} and direct reconstruction of heavy flavour mesons \cite{ALICE:2012ab, Abelev:2013lca, Abelev:2014ipa} have shown both substantial suppression in the yield of heavy quarks due to energy loss and significant azimuthal anisotropy. 


This \papertype\ presents ATLAS measurements of heavy flavour muon production measured over $|\eta| {<} 1$\footnote{ATLAS uses a right-handed coordinate system
  with its origin at the nominal interaction point (IP) in the center
  of the detector and the $z$-axis along the beam pipe. The $x$-axis
  points from the IP to the center of the LHC ring, and the $y$-axis
  points upward.  Cylindrical coordinates $(r,\phi)$ are used in the
  transverse plane, $\phi$ being the azimuthal angle around the beam
  pipe. The pseudorapidity is defined in terms of the polar angle
  pipe. The pseudorapidity is defined in terms of the polar angle
  $\theta$ as $\eta=-\ln\tan(\theta/2)$.} 
using 0.14~nb$^{-1}$ of \PbPb data at $\sqn = 2.76$~TeV and 4.0~pb$^{-1}$ of \pp data at $\sqs = 2.76$~TeV, collected during LHC operation in 2011 and 2013, respectively. The measurements are performed for several intervals of collision centrality, characterized using the total transverse energy measured in the forward calorimeters, and for different \pT intervals spanning the range $4 {<} \pT {<} 14$~GeV. 

The heavy flavour muon differential per-event yields in \PbPb collisions and differential cross-sections in \pp collisions are used to calculate the heavy flavour muon \RAA as a function of \pT in different \PbPb centrality intervals. In addition, the heavy flavour muon \vt is measured as a function of \pT and collision centrality using the event-plane method with the second-order event plane angle, $\Psi_2$, measured in the forward calorimeters.

Heavy flavour muons are statistically separated from background muons resulting from pion and kaon decays and hadronic interactions using a variable that compares the momenta of the muons measured in the inner detector and muon spectrometer. Over the \pT range of the measurement, the residual irreducible contamination of non-heavy flavour muons to the measurement, is ${\lesssim} 1\%$  including contributions from \Jpsi decays \cite{Aad:2011rr}.

\section{Data analysis}
\label{sec:analysis}

The measurements presented in this \papertype\ are obtained using the ATLAS muon spectrometer (MS), inner detector (ID), calorimeter, trigger and data acquisition systems. A detailed description of these detectors and their performance in \pp collisions can be found in  Ref.~\cite{Aad:2008zzm}.
The \PbPb events selected for this analysis are required to have a reconstructed vertex and a time difference between the signals in two
Minimum Bias Trigger Scintillator detectors of less than 5~ns. At least one reconstructed collision vertex is required  in \pp collisions. The centrality of \PbPb collisions is characterised by transverse energy measured in the ATLAS forward calorimeter (FCal). Results presented in this \papertype, are worked out for several intervals 
ordered from the most central to the most peripheral collisions: 0--10\%, 10--20\%, 20--30\%, 30--40\%, and 40--60\%. 

Muons used in this analysis are obtained by combining tracks reconstructed in the MS with the tracks measured in the ID. The tracks are required  to have momentum $p{>}3$~(1.2)~GeV in the ID and the MS respectively and to satisfy criteria on the number of hits in ID that are the same for the \pp and \PbPb data. Transverse and longitudinal impact parameters of the track with respect to a reconstructed event vertex are required to be less than 5~mm. 
Muons selected for this analysis are constrained at low \pT to be above 4~GeV by the dependence of the muon trigger and reconstruction efficiencies while at high \pT by the available statistics of the \PbPb data. The  muon $\eta$ interval is chosen for optimal muon performance. A total of 9.2~million  (1.8~million) muons are reconstructed within these kinematic ranges from 8.7 million (1.8 million) events recorded using the \PbPb  (\pp) muon triggers.

The performance of the ATLAS detector and offline analysis in measuring muons is evaluated using Monte Carlo samples obtained from \textsc{Geant4}-simulated~\cite{Agostinelli:2002hh} $\sqs = 2.76$~TeV \pp dijet events produced with the \textsc{Pythia} event generator ~\cite{Sjostrand:2006za} (version 6.423 with parameters chosen according to the AUET2B tune ~\cite{mctunes}). To account for the large occupancy in the detector readout channels in central \PbPb collisions \textsc{Pythia} events were overlayed onto \PbPb collision events selected with the minimum-bias trigger.
For both the \pp and \PbPb measurements, the muon reconstruction efficiency increases by about 15\% between $\pT{=}4$ and 6~GeV 
above which it is approximately constant at $\sim$0.87 and $\sim$0.84 for the \pp and \PbPb data, respectively. The \PbPb reconstruction efficiency is independent of centrality within uncertainties. The \PbPb muon trigger efficiency increases from 0.60 at $\pT = 4$~GeV to $\sim$0.75 at 6~GeV, above which it is approximately constant. The \pp muon trigger efficiency increases from 0.52 for $4 {<} \pT {<} 4.5$~GeV to 0.82 for $\pT {>} 5.5$~GeV, above which it remains constant. 

\begin{figure}[!htb]
\centerline{
\includegraphics[width=0.9\textwidth]{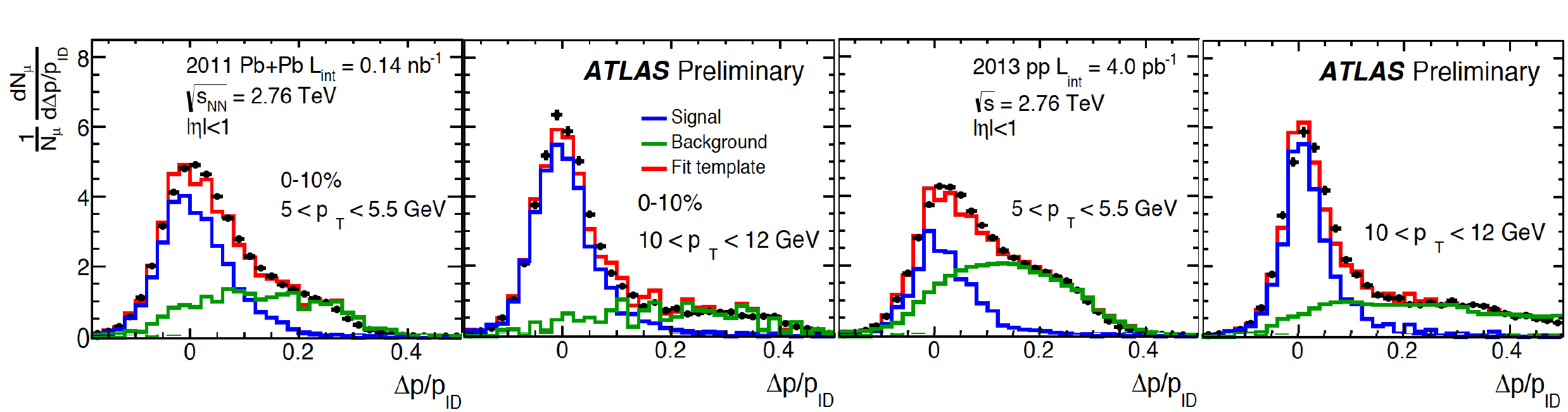}%
}
\caption{
Results of template fits to the 0--10\% most central \PbPb collision data (left two panels) and to the \pp collision data (right two panels) for two \pT ranges indicated in the panels. The black points represent the data, blue and green lines represent the signal and background template distributions and the red lines represent the combined template distributions.
Figure from~\cite{ATLAS-CONF-2015-053}.
} 
\label{fig:ploss}
\end{figure}

The muons measured in the \pp and \PbPb data sets contain background from in-flight decays of pions and kaons, muons produced from the decays of particles produced in hadronic showers, and mis-associations of ID and MS tracks. Previous studies have shown that the signal and background contributions to the reconstructed muon sample can be discriminated statistically \cite{Aad:2011rr}. This analysis relies on the fractional momentum imbalance, $\Deltaploss/\pID$, shown in Fig.~\ref{fig:ploss} which quantifies the difference between the ID and MS measurements of the muon momentum after accounting for the energy loss of the muon in the calorimeters. 

\section{Results}
\label{sec:results}
The heavy flavour muon differential cross-section in \pp and differential yields in \PbPb collisions are determined from the template fit procedure shown in Fig.~\ref{fig:ploss}. 
 \begin{figure}[htb!]
\centerline{
\includegraphics[width=0.33\textwidth]{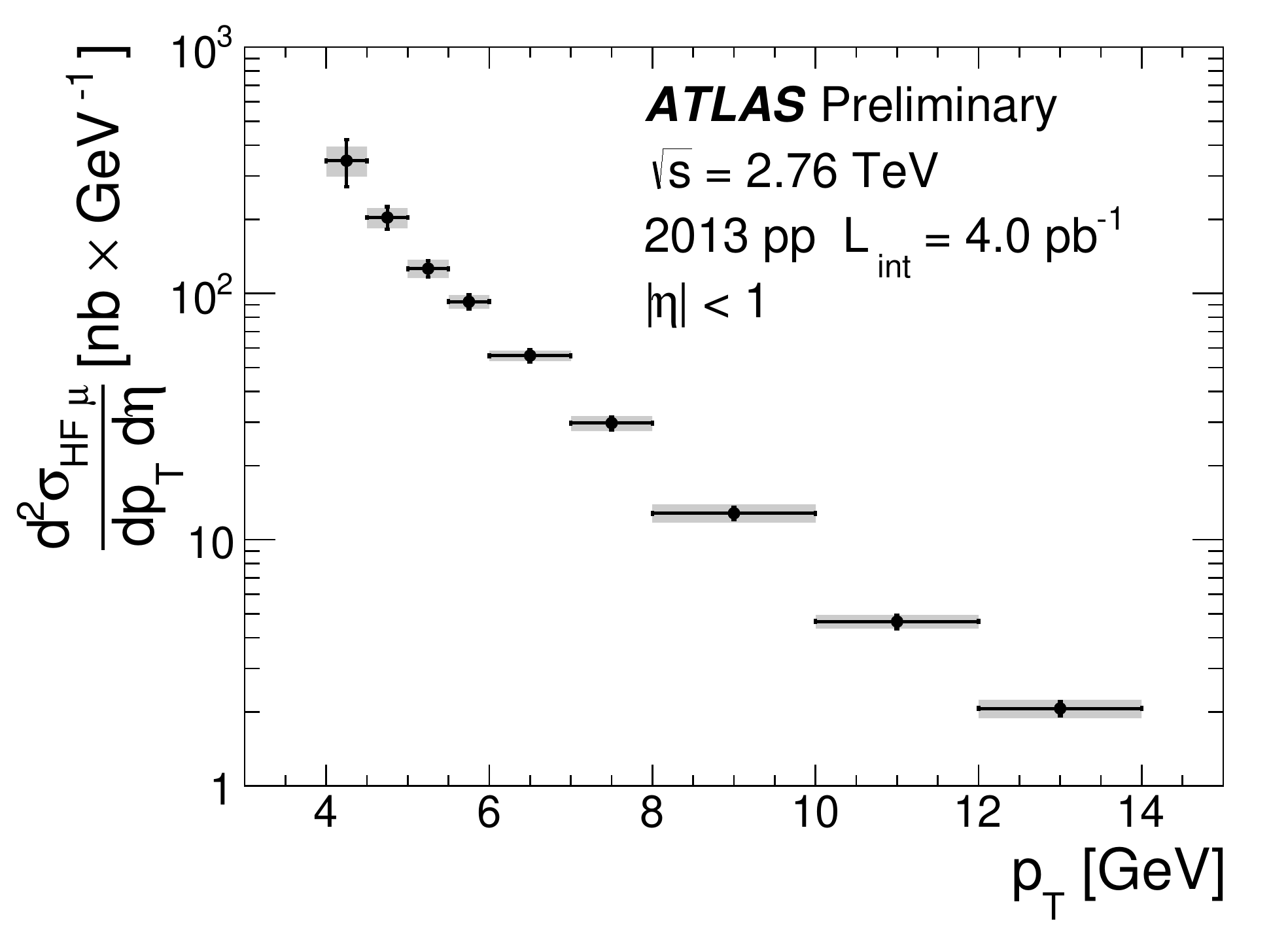}
\includegraphics[width=0.33\textwidth]{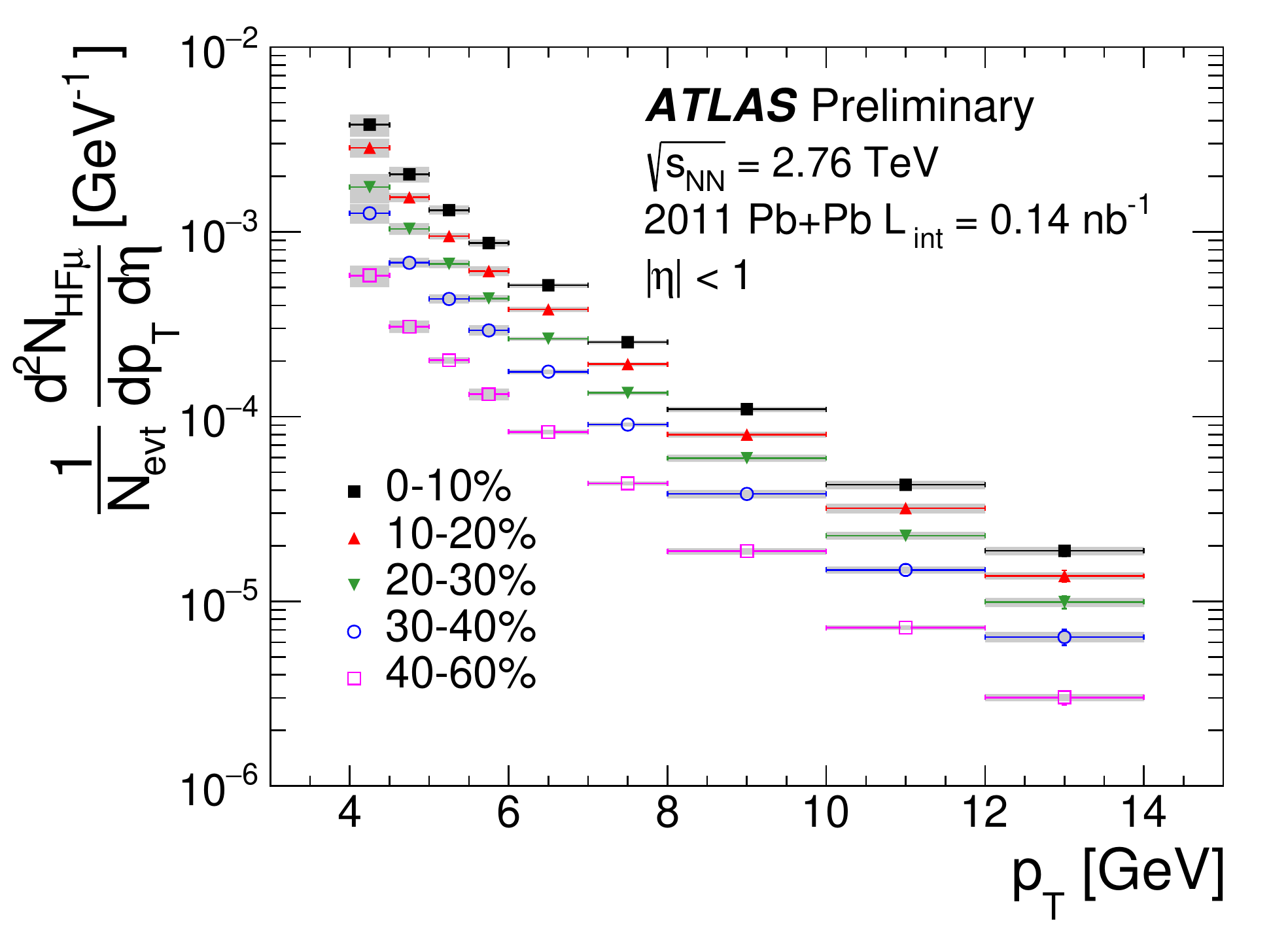}
\includegraphics[width=0.32\textwidth]{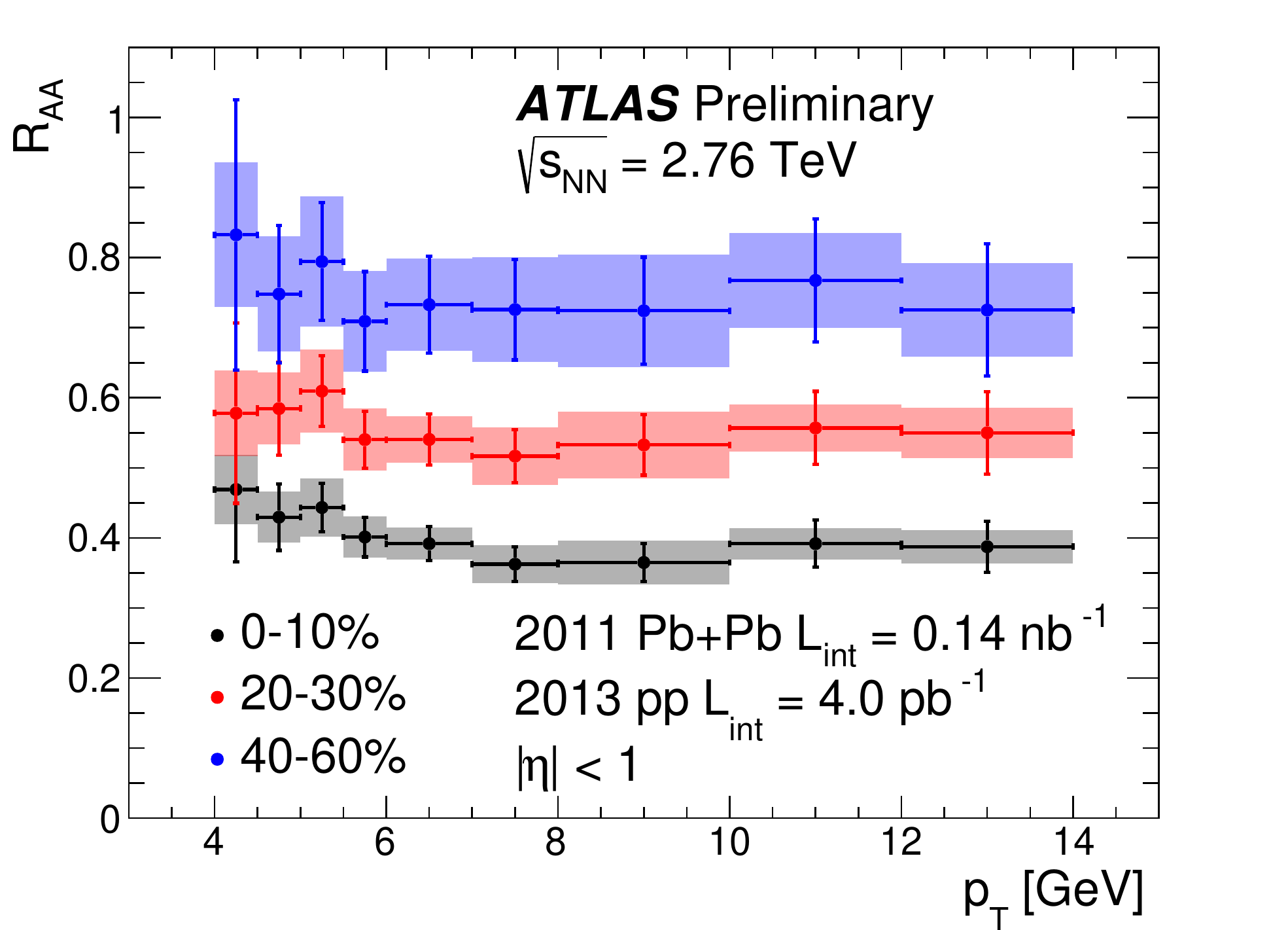}
}
\caption{
Left: \pp heavy flavour muon differential cross-section. Middle:  \PbPb  heavy flavour muon differential per-event yields. Right: heavy flavour muon \RAA measured as a function of \pT. The error bars represent statistical uncertainties on the data, while the systematic uncertainties, including the contribution from luminosity, are indicated by the shaded bands. Figure from~\cite{ATLAS-CONF-2015-053}.
} 
\label{fig:ppsig}
\end{figure}
The measurement of the heavy flavour muon differential cross-sections and per-event yields are subject to systematic uncertainties arising from the muon trigger selection, muon reconstruction efficiencies, the template fitting procedure, muon \pT resolution, and the \pp luminosity explained in Ref.~\cite{ATLAS-CONF-2015-053}. The results for \pp and \PbPb are shown in Fig.~\ref{fig:ppsig} in the left and middle panels respectively. The right panel of the figure shows heavy-flavour muon \RAA, which decreases between peripheral (40--60\%) and more central collisions reaching a value of $\sim$0.4 in the 0--10\% centrality interval. The measured \RAA appear to be \pT-independent within the uncertainties of the measurement. These results are consistent with previous results from the ALICE experiment~\cite{Abelev:2012qh}, but have much smaller uncertainties. 

\begin{figure}[!htb]
\centerline{
\includegraphics[width=0.40\textwidth]{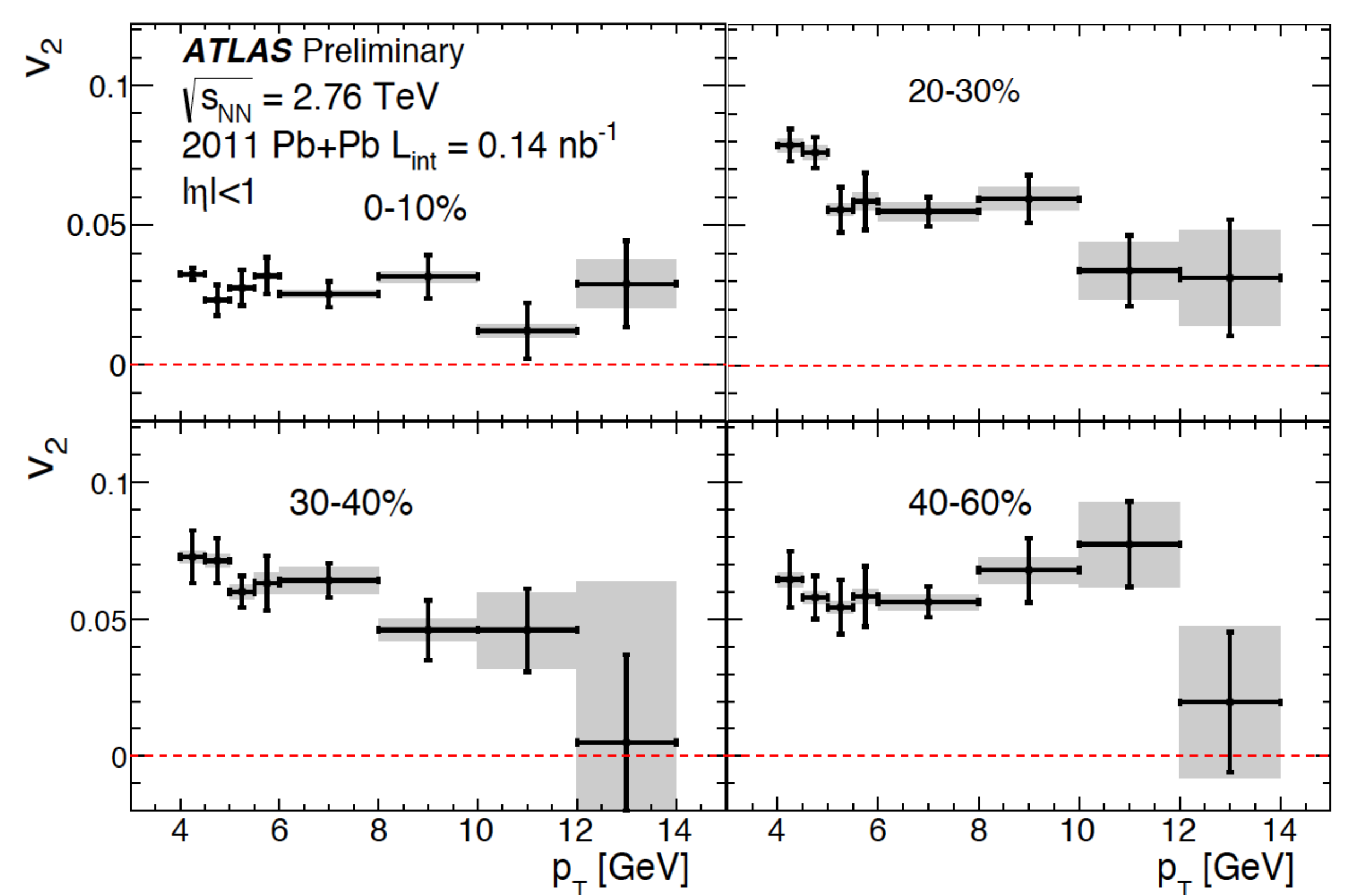}%
\includegraphics[width=0.36\textwidth]{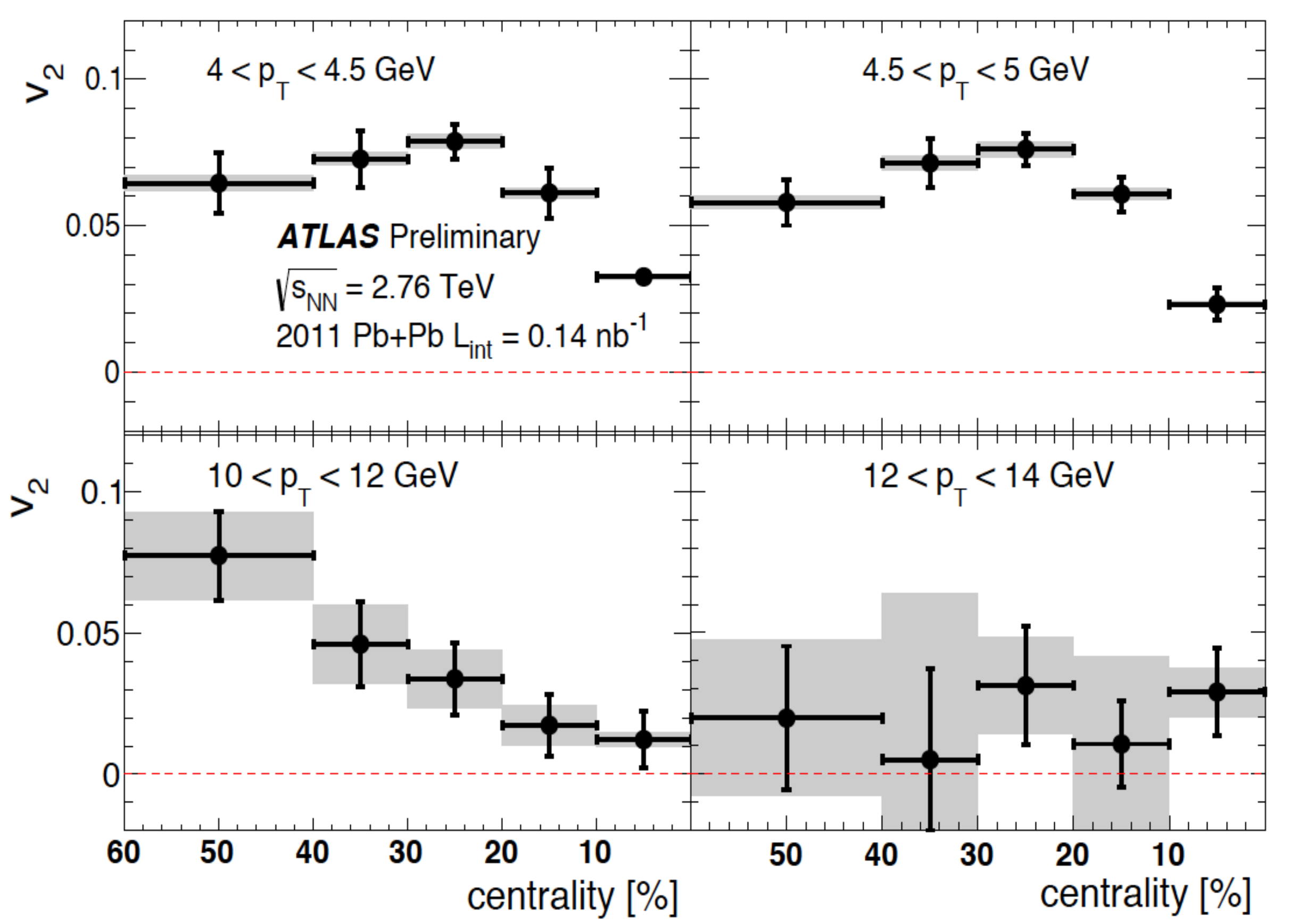}%
}
\caption{
The \pT dependence (left) and centrality dependence (right) of the heavy flavour muon \vt. The error bars and shaded bands represent statistical and systematic uncertainties respectively. Figure from~\cite{ATLAS-CONF-2015-053}
} 
\label{fig:v2_pt_dep}
\end{figure}

The heavy flavour muon \vt values are measured by evaluating the yields differentially with respect to the event plane. 
The extracted \vt values, are then corrected to account for the event plane resolution. The sources of the systematic uncertainties in the \vt measurements are mostly the same as those in the \RAA measurements, however, the trigger and tracking efficiencies do not have a significant effect on the \vt. The resolution-corrected \vt values are plotted in the left panel of Fig.~\ref{fig:v2_pt_dep} as a function of \pT for several centrality intervals used in this analysis. 

Over the 20--40\% centrality range, the \vt is largest at the lowest measured \pT of 4~GeV and decreases for higher \pT. However, in the 0--10\% and 40--60\% centrality intervals, no clear \pT dependence is visible. For all centralities, significant non-zero \vt is observed even at \pT of 10~GeV. Right panels of the figure show the same set of results plotted as a function of centrality for different \pT intervals. For low \pT range the centrality dependence of the heavy flavour muon \vt is qualitatively similar in shape, but considerably smaller in magnitude, to that for charged hadrons of similar \pT~\cite{ATLAS:2011ah,ATLAS:2012at}. In this \pT range, the \vt first increases from central to mid-central events, reaches a  maximum between 20--40\% centrality, and then decreases. Over the \pT range of 8--12~GeV 
the \vt increases monotonically from central to peripheral events, However, the associated statistical and systematic uncertainties are considerably larger. This monotonically increasing centrality dependence of the \vt  at high \pT is also seen in the inclusive charged hadron \vt~\cite{ATLAS:2011ah, ATLAS:2012at}. For the highest \pT interval of $12{<}\pT{<}14$~GeV, the statistical and systematic errors are too large to clearly understand the centrality dependence of the \vt.

\bibliographystyle{elsarticle-num}
\bibliography{qm15_atlas_muons_milov}







\end{document}